\documentclass[%
 reprint,
showpacs,preprintnumbers,
 amsmath,amssymb,
pra,
]{revtex4-1}
\usepackage[utf8]{inputenc}	
\usepackage[T1]{fontenc}	
\usepackage{graphicx}		
\usepackage{dcolumn}		
\usepackage{booktabs}		
\usepackage{mathtools}		
\usepackage{amsmath,array}
\usepackage{relsize}
\usepackage{enumerate}
\usepackage[squaren]{SIunits}	
\usepackage {natbib}

\let\originaleqref\eqref
\renewcommand\figurename{Fig.}
\renewcommand{\eqref}{Eq.~\originaleqref}

\newcommand{\fref}[1]{\figurename~\ref{#1}}
\newcommand{\lref}[1]{\eqref{#1}}

\newcommand{\ket}[1]{\left| #1 \right>} 
\newcommand{\bra}[1]{\left< #1 \right|} 

\newcommand{\ttt}{(\tau,\theta)} 

\newcommand{\pdiff}[2]{\frac{\partial #1}{\partial #2}}
\newcommand{\s}{\sigma}

\graphicspath{{Billeder_reviewed/}}

\begin{document}
\title{Estimation of atomic interaction parameters by photon counting}
\author{Alexander Holm Kiilerich and Klaus Mølmer}
\affiliation{Department of Physics and Astronomy, Aarhus University, Ny Munkegade 120, DK 8000 Aarhus C. Denmark}
\date{\today}
\bigskip

\begin{abstract} 
Detection of radiation signals is at the heart of precision metrology and sensing. In this article we show how the fluctuations in photon counting signals can be exploited to optimally extract information about the physical parameters that govern the dynamics of the emitter. For a simple two-level emitter subject to photon counting, we show that the Fisher information and the Cram\'er-Rao sensitivity bound based on the full detection record can be evaluated from the waiting time distribution in the fluorescence signal which can, in turn, be calculated for both perfect and imperfect detectors by a quantum trajectory analysis. We provide an optimal estimator achieving that bound.
\end{abstract}

\maketitle
\noindent
\section{introduction}
Atoms and molecules find wide applications in fundamental tests of physics and as field and inertial sensors. Their quantized energy levels permit their use as clocks and frequency references, while the special role of measurements in quantum mechanics imposes fundamental sensitivity limits. In a quantum measurement the outcome is governed by probabilistic rules, and the optimal estimation of physical parameters by measurement data becomes a statistical inference problem.  In this article, we address the problem of continuous quantum measurements and the optimal use of full measurement records in parameter estimation. The fluorescence signal from a laser excited two-level atom has a mean intensity, which in steady state is a known function of the field-atom detuning, the Rabi-frequency and the atomic decay rate - but the full record of photon count events contains much more information than the mean signal. In the quantum trajectory picture of resonance fluorescence \cite{CARMICHAEL, MCWF, WM}, each detector click is accompanied by an atomic quantum jump into the ground state and a subsequent transient evolution which leads to a modulation of the probability distribution for subsequent detection events. The time record of intervals between detector clicks or the cumulants of the counting statistics \cite{bruderer}, therefore allow better discrimination between different values of the Rabi-frequency than the steady state mean intensity.

In recent works \cite{likelihood,Mabuchi1996, PhysRevA.64.042105}, the derivation of likelihood functions by application of Bayes' rule to continuous measurement records has been shown to follow in a quite straightforward manner from the theory of stochastic master equations and quantum trajectories.
In \cite{likelihood} a numerical procedure to achieve this goal, through simulation of the stochastic master equation, was presented with explicit attention to the cases of homodyne/heterodyne detection and photon counting. The expected asymptotic accomplishment of parameter estimation by stochastic measurement records was furthermore addressed in \cite{likelihood}, and it was shown that Monte Carlo simulations of quantum trajectory ensembles can be used to estimate the Fisher information and Cram\'{e}r Rao bound \cite{Cramer} associated with any particular detection scheme. For numerical examples of the Fisher information in the case of photon counting, obtained by such simulations, see \cite{likelihood}, and for a comparison between the Fisher information and Cram\'{e}r Rao bound for photon counting and homodyne detection (as well as a theoretical optimal limit of sensitivity), see  \cite{Fisher_open}.

Due to the combined action of the Hamiltonian and measurement back action on the system evolution, measurement records obtained in experiments generally depend on the system parameters in a complicated manner, and one should not expect any analytical expression for the Fisher information connected to a particular detection scheme. In this article, however, we exploit a specific property of the quantum trajectories associated with fluorescence records from two-level emitters: The signal is a discrete set of detection times, and after each detection event the system populates the same state and recommences the same evolution. This implies that the distribution of delay times between detector clicks yields the same information as the full detection record. Since this distribution can be determined analytically,
our method permits analytic calculations and interpretation of the results in different limits along with a straightforward incorporation of finite detector efficiency in the analysis.

The article is organized as follows. In Sec. \ref{sec:sim} we review the connection between stochastic quantum measurement dynamics and the Bayesian parameter estimation problem. In Sec. \ref{sec:stat} we present the specific statistical analysis of the Fisher information based on delay times between detector clicks in photon counting experiments. In Sec. \ref{sec:the delay function} we provide a quantum optical analysis of the delay distribution function, valid for perfect and imperfect detectors.
In Sec. \ref{sec:conclusion} we conclude and present an outlook.

\section{Quantum trajectories and Bayesian parameter estimation}
\label{sec:sim}
A quantum measurement process can be regarded as a filter that allows particular components of the system wave function or density matrix to pass as representative of the state of the system if they are in accordance with the outcome of the measurement. Such a quantum filter is readily augmented to include the identification of unknown classical parameters by treating them as physical properties of ancillary system degrees of freedom. A constant, unknown parameter can thus be formally described in the same way as a quantum non-demolition (QND) variable, \cite{Braginsky01081980}, of an ancillary quantum system, and the evolution of the joint system by quantum measurement theory may effectively accomplish a QND measurement of that parameter \cite{PhysRevA.70.052102}. An equivalent approach assigns possible candidate values to a \textit{particle filter}, where separately evolved quantum states, assuming different parameter values, carry weight factors which are changed conditioned on the measurement outcomes \cite{PhysRevLett.108.230401,PhysRevA.79.022314}.

In this section we recall the quantum jump dynamics of a laser driven two-level atom subject to fluorescence detection, and we present examples of how the observation of very few photons serves to distinguish between different values of the unknown Rabi-frequency.

\subsection{Photon counting from a laser driven two-level atom}
\label{section:photon_counting}
The master equation for the density matrix $\rho_t$ of an atom with ground state $|g\rangle$ and excited state $|e\rangle$, coupled to a laser field can in the rotating wave approximation be written ($\hbar=1$):
\begin{align}
\mathrm{d}{\rho}_t=-i[\hat{H}_0,{\rho}_t]\mathrm{d}t-\frac{\Gamma}{2}\lbrace\hat{\sigma}^{\dagger}\hat{\sigma},{\rho}_t\rbrace\mathrm{d}t + \Gamma\hat{\sigma}{\rho}_t\hat{\sigma}^{\dagger}\mathrm{d}t,
\label{masterligning_middel}
\end{align}
where $\hat{\s}=\ket{g}\bra{e}$ and $\Gamma$ is the excited state decay rate. In the frame rotating with the frequency of a monochromatic laser beam, the Hamiltonian is
\begin{align}
\hat{H}_0=-\delta\hat{\s}^{\dagger}\hat{\s}+\frac{\Omega}{2}(\hat{\s}^\dagger+\hat{\s}),
\label{hamilton2}
\end{align}
where $\Omega$ is the Rabi frequency and $\delta$ is the laser-atom detuning.

Eq. (\ref{masterligning_middel}) can be unravelled into stochastic evolution corresponding to random measurement back action on the atom due to the detection of the emitted radiation.
The last term in the master equation (\ref{masterligning_middel}) is related to the process of photon emission, yielding an incoherent increase in the ground state population. The other terms in (\ref{masterligning_middel}) are associated with the atomic state component in the absence of emitted photons. Monitoring the fluorescence signal conditions the state evolution on the resulting measurement record. Photon detection causes a jump to the ground state $\rho_t \rightarrow \Gamma \hat{\s}\rho_t \hat{\s}^\dagger \mathrm{d}t \propto |g\rangle\langle g|$, while the effect of detecting no photon in the field is to merely omit the contribution from the last term in (\ref{masterligning_middel}).
The probabilities for these events to occur are given by the trace of the respective terms, which yields the expected probability $\Gamma \rho_{ee} \mathrm{d}t$ for jumping into the ground state and $1-\Gamma \rho_{ee} \mathrm{d}t$ for continuous no-jump evolution during time intervals with no photon detection \cite{MCWF}.

The field is reset to the vacuum state, the atomic state is renormalized and the evolution proceeds after the projection of the system on the one- and zero-photon components. An initially pure state remains pure, i.e., it can be described by a state vector during this dynamics, and a wave function simulation scheme, where one uses random numbers to synthesize typical detection records, averaged over many such evolutions, yields the same results as the master equation \eqref{masterligning_middel}, \cite{MCWF}. The upper panel in \fref{fig:likelihood_3} shows the result of a single simulation carried out for a two-level atom with a Rabi frequency $\Omega_0=5\Gamma$, and a vanishing detuning $\delta=0$. The figure shows oscillations in the excited state population $|c_e|^2$ interrupted by quantum jumps resetting the dynamics.
In the simulated time span of $40 \Gamma^{-1}$, 22 jumps are registered, consistent with an average excited state population of  $\sim 0.5$. While the number of clicks is statistically compatible with a rather wide range of values of $\Omega$, the time intervals between detector clicks provide sharper information about $\Omega_0$ than the total number of clicks.

\subsection{Parameter estimation by Bayes' rule}
The no-jump and jump dynamics are associated with detection of photons, and  the probabilities of the respective detection events depend on the solution of the stochastic dynamics, which in turn depends on the parameters in the master equation. The measurement process therefore continuously provides information about these parameters.
The formal treatment of this acquisition of information follows Bayes' rule: The probability for an unknown parameter to have a given value $\theta$, conditioned on the stochastic measurement outcome $D$, is given by the probability for that outcome conditioned on the value $\theta$ together with their unconditional (prior) probabilities,
\begin{align}
P(\theta|D)=\frac{P(D|\theta)P(\theta)}{P(D)}.
\label{bayes rule}
\end{align}
For our application, $D$ is the click or absence of a click in a time interval $\mathrm{d}t$. The click, e.g., occurs with a probability $\Gamma |c_e|^2 \mathrm{d}t$, conditioned on the value of the coupling strength in the master equation through its influence on the excited state population. \eqref{bayes rule} hence yields an update of the probability $P(\theta)$, assigned to different values of $\theta$ prior to detection in the time interval $\mathrm{d}t$. The update rule is applied at every time step and thus acts as a filter on the values of the parameter $\theta$.

\begin{figure}
\centering
\includegraphics[width=1\columnwidth]{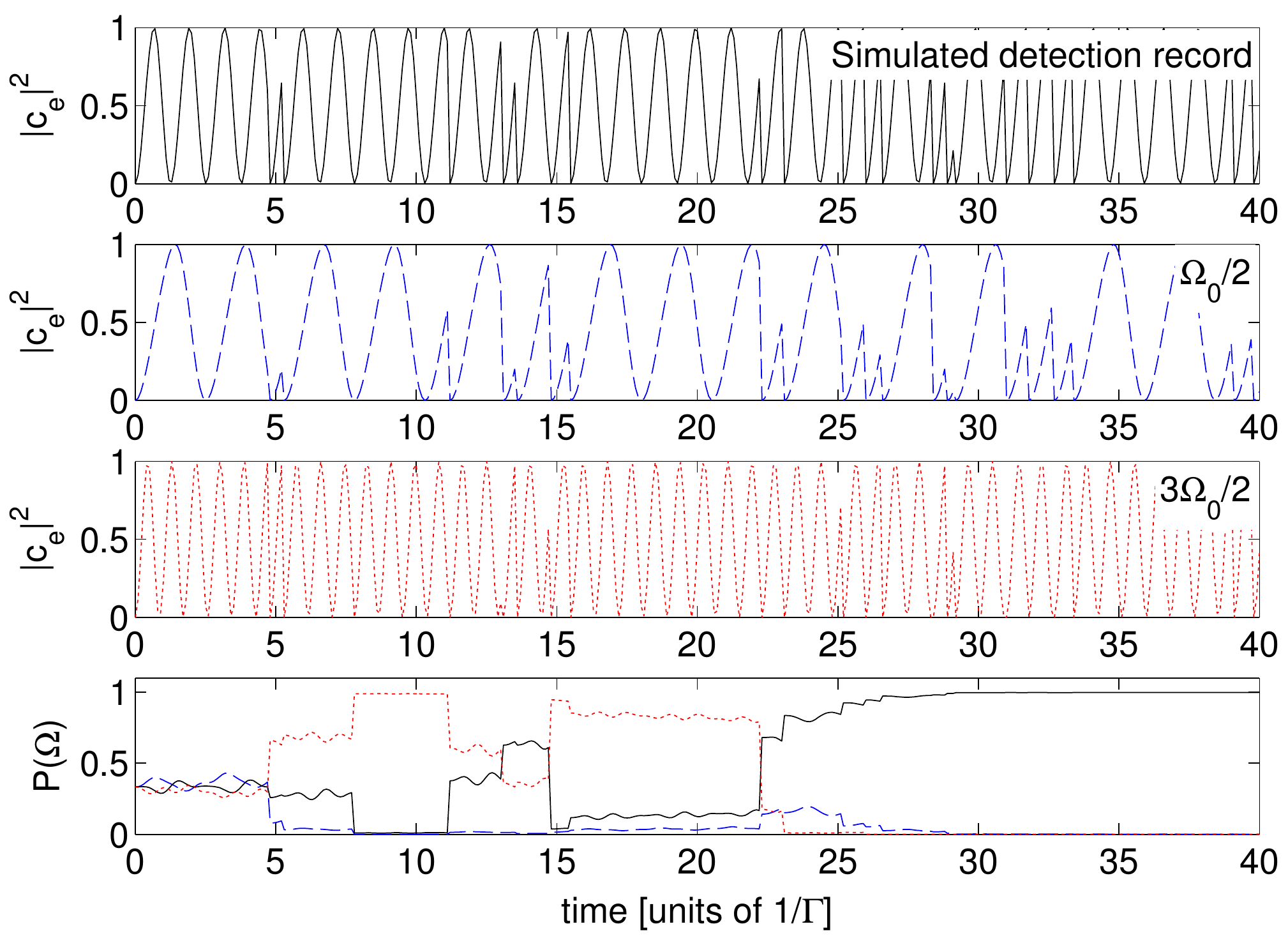}
\caption{\textsl{(Color online) The upper panel shows the excited state population resulting from a quantum trajectory simulation with $\Omega_0=5\Gamma$ and $\delta=0$. The two middle panels show the evolution conditioned on the same detection events as in the upper panel, but assuming $\Omega_0/2$ and $3\Omega_0/2$, respectively. The lower panel shows the evolution of the probabilities $P(\Omega)$ for the candidate values, $\Omega=\Omega_0$ (solid, black, upper curve at $t=25\Gamma^{-1}$), $\Omega=\Omega_0/2$ (dashed, blue, middle curve at $t=25\Gamma^{-1}$) and $\Omega=3\Omega_0/2$ (dotted, red, lower curve at $t=25\Gamma^{-1}$ ), conditioned on the measurement record.}}
\label{fig:likelihood_3} 
\end{figure}

Note that the denominator in \eqref{bayes rule} is independent of $\theta$, and it is in practical calculations not necessary to compute it at every step of the update protocol. One may, thus, describe the knowledge about the variable $\theta$ by (un-normalized) likelihood functions, $L(\theta)$ and $L(\theta|D)$,  which are merely proportional to the probability distribution. The final conditional likelihood function is given by simply multiplying together all the probability factors assigned by quantum measurement theory to the count and no-count events \textit{actually occurring} during the experiment \cite{likelihood}.
We illustrate this principle in \fref{fig:likelihood_3} by propagating stochastic wave functions assuming two alternative values, $\Omega_0/2$ and $3\Omega_0/2$, for the Rabi frequency.
The results in the second and third panel reveal slower and faster Rabi oscillations between the jumps, which are selected in conformity with the (simulated) experimentally observed detector clicks, i.e. according to the dynamics governed by the true value $\Omega_0=5\Gamma$ of the Rabi frequency. Consequently in both the second and third panel we observe jumps at times, where the excited state population is small. Such jumps are not very probable, and through Bayes rule the conditional probabilities for the values $\Omega_0/2$ and $3\Omega_0/2$ are correspondingly suppressed. Assuming that only $\Omega_0/2$,  $\Omega_0$, and $3\Omega_0/2$ are allowed values for the Rabi frequency, and that they are a priori equally probable, panel 4 in Fig.(\ref{fig:likelihood_3}) shows the evolution of the respective probabilities as the detection record unfolds.

Let us go into detail with some of the features of the plot.
One notices how non-click periods lead to smooth, continuous evolution while the clicks demand more pronounced changes. The first jump is in conflict with a Rabi frequency of $\Omega_0/2$, while it favors $3\Omega_0/2$ which has the largest excited state amplitude here. The probability for $\Omega=3\Omega_0/2$, indeed, dominates for a while until further detection events signify that the clicks are actually selected according to the evolution with the correct value $\Omega_0$. The true value acquires a probability close to unity after merely 20 detector clicks.

\begin{figure}
\centering
\includegraphics[width=1.0\columnwidth]{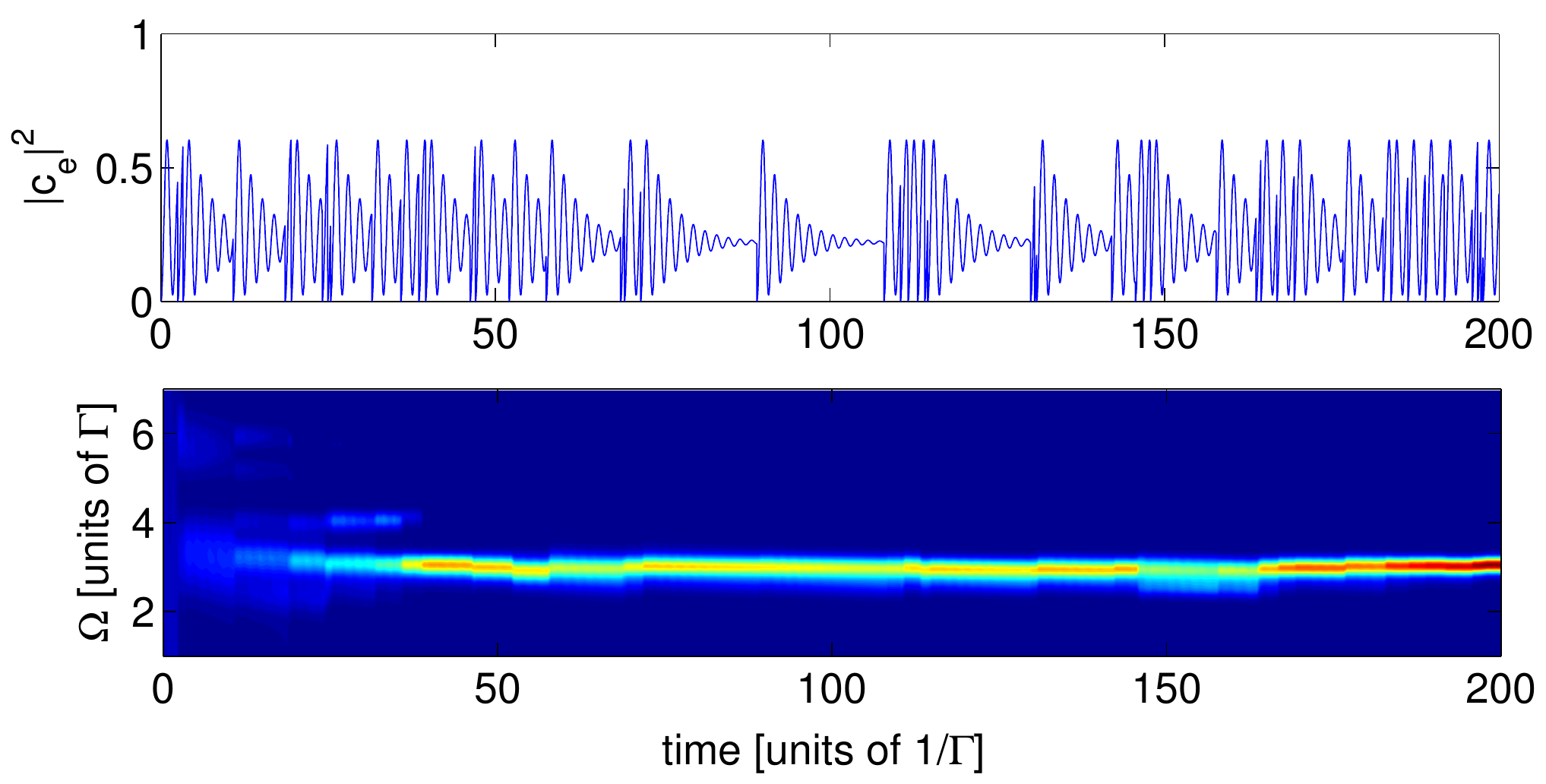}
\caption{\textsl{(Color online) The upper panel illustrates a simulated data record assuming $\Omega_0=3\Gamma$ and $\delta=2\Gamma$. The lower panel shows the resulting evolution of a quasi continuous probability distribution for the Rabi frequency $\Omega$.
The total number of detection events is 51.}}
\label{fig:likelihood_kontinuert} 
\end{figure}

\fref{fig:likelihood_kontinuert} shows the results of applying the same procedure for a wider range of candidate values of the Rabi frequency on a fine grid and for longer time. The upper panel shows the excited state population associated with the simulation of a detection record, and the color plot shows how the likelihood function gradually develops a narrow peak around the correct value of the Rabi frequency.

\section{Fisher information}
\label{sec:stat}
In the previous section we presented Bayesian inference as an incremental operation
utilizing at every time step the most recent likelihood function as the prior and the current measurement result to yield the new, conditioned likelihood function. To address the asymptotic behavior of the uncertainty in our parameter estimate we will now apply Bayes' rule \eqref{bayes rule} to the case, where $D$ denotes the entire detection record. Hence we regard the final likelihood function for the unknown parameter $L(\theta|D)$ as the updated probability, conditioned on all available data, and we address the asymptotic behavior of the uncertainty on our parameter estimate as a function of the total number of detected photons.

\subsection{Fisher information of measurement records}

According to the Cram\'{e}r-Rao Bound (CRB) \cite{Cramer}, any unbiased estimator $S(\theta)$ for an unknown parameter $\theta$, determined by $K$ independent measurements, fluctuates around the actual value with a statistical variance
\begin{align}
\left(\Delta S(\theta)\right)^2 \geq\frac{1}{KF(\theta)},
\label{CRB}
\end{align}
where
\begin{align}
F(\theta)=-\sum_D \frac{\partial^2 \ln L(D|\theta)}{\partial \theta^2}L(D|\theta) \
\label{fisher kompleks}
\end{align}
is the so-called Fisher information, \cite{Cramer,likelihood}.

Rather than $K$ independent measurements, we have a single experiment in mind, where one observes the fluorescence emitted by only one atom for a long time. In this case, however, the meaning of the Cram\'{e}r-Rao Bound as an asymptotic limit still holds: we obtain in a single data record ($K=1$) the equivalent of a large number of independent measurement outcomes, and rather than the explicit $K$ factor multiplying $F(\theta)$ in \eqref{CRB}, the Fisher information itself becomes proportional to $N$, the total number of photons detected during the accumulation of data.

\subsection{Fisher information and waiting time distributions}

The expression for the Fisher information \eqref{fisher kompleks} makes reference to the variation of the likelihood function over the set of possible measurement records $D$. The weighted summation (integral) over all data records makes the direct evaluation of \eqref{fisher kompleks} a formidable task. In this section, we will present an alternative description that provides a simple evaluation of  the Fisher information for two-level quantum jump dynamics.

\begin{figure}
\centering
\includegraphics[width=1\columnwidth]{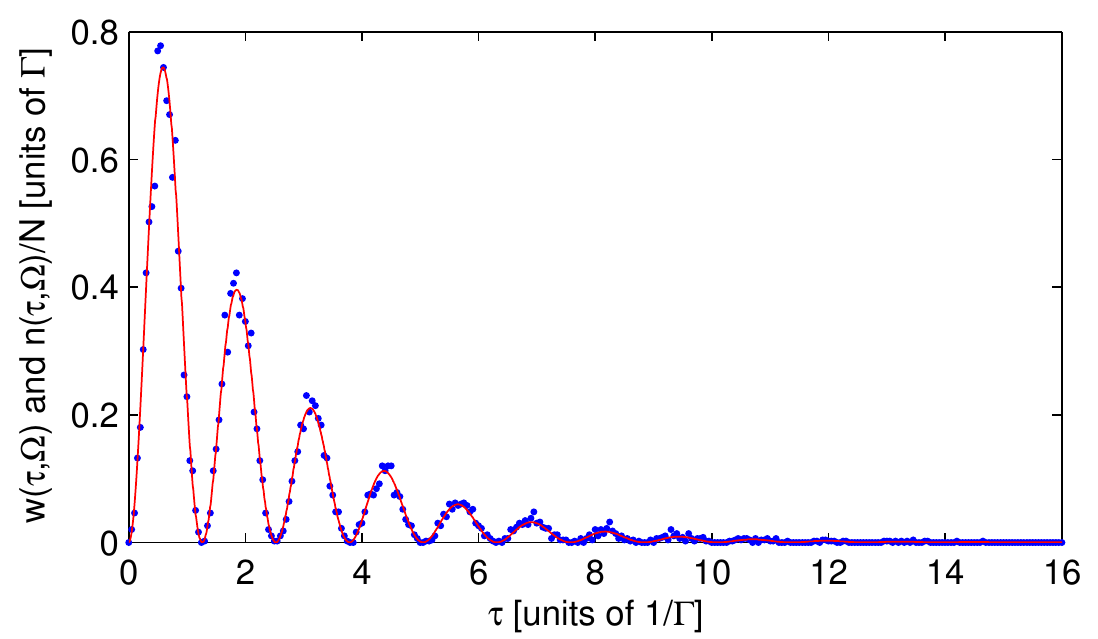}
\caption{\textsl {(Color online) The blue noisy dots shows the distribution from a simulated data record of 10000 detection events. The parameters are $\delta=0$ and $\Omega=5\Gamma$. The red curve is the corresponding theoretical waiting time distribution $w(\tau,\theta)$ for time intervals between detector clicks.}}
\label{fig:delay_test} 
\end{figure}

Our theory relies on two observations: (\textit{i}) the full data record is unambiguously represented by a list of the instants of time $t_i$ where a photon is detected, (\textit{ii}) the atom jumps to the same state after each detection event, and the waiting times, i.e., the time intervals $\tau_i = t_i-t_{i-1}$ between subsequent jump events are therefore uncorrelated stochastic variables with the same probability distribution. It follows from (\textit{ii}) that there is no information in the actual order of the different intervals recorded. Thus, the only relevant information in the detection record is the distribution of registered intervals between jumps. In \fref{fig:delay_test} we show the distribution of 10000 time intervals between simulated quantum jumps (blue dots). The comparison with the theoretical waiting time distribution (red curve) constitutes the basis of the parameter estimation since a higher or smaller value of the Rabi frequency would change the oscillation period in the distribution of  waiting times.

The data record $D$ is without loss of information reduced to the registered distribution of waiting times between jumps. After the detection of a total number of $N$ photons, the number $k$ of occurrences of waiting times in a small interval $[\tau,\tau + d\tau]$ follows Poisson statistics, $P(k|\theta) = \frac{(\overline{n}(\tau,\theta)\mathrm{d}\tau)^{k}}{k!}e^{-\overline{n}(\tau,\theta)\mathrm{d}\tau}$ where the mean value is $\overline{n}(\tau,\theta)d\tau=Nw(\tau,\theta) \mathrm{d}\tau$, and $w(\tau,\theta)$ is the normalized waiting time distribution. The Fisher information associated with a counting signal sampled from uncorrelated Poisson distributions plays a role, e.g., in microscopy where position dependent signals of scattered coherent light are used to track the position of scatterers with high resolution \cite{CRB,fischer}. The Fisher information in that problem is known, and translating the position argument to the waiting time argument $\tau$, we conclude that the sensitivity of sampling is governed by Eq.(\ref{CRB}), where the Fisher information reads \cite{CRB}
\begin{align}
F(\theta)=\int \frac{1}{\bar{n}\ttt}\left(\frac{\partial\bar{n}\ttt}{\partial\theta}\right)^2 \, \mathrm{d}\tau.
\label{specialcase}
\end{align}
If $\overline{n}(\tau,\theta)$ is known, only a one-dimensional integral has to be computed - a trivial task in comparison with the calculation envisaged in \lref{fisher kompleks}.

We have $N=\int \overline{n}(\tau,\theta)d\tau$ and, hence, we see directly that the Fisher information is proportional to the total number of detected photons,
\begin{align}
F(\theta)&=\frac{N}{a^2}.
\label{fischer delay}
\end{align}
The constant of proportionality is given by \eqref{specialcase}, and can also be written \cite{CRB}
\begin{align}
\frac{1}{a(\theta)^2}=4\int \left(\frac{\partial \Phi\ttt}{\partial\theta}\right)^2 \, \mathrm{d}\tau
\label{a}
\end{align}
where $\Phi\ttt=|\sqrt{w\ttt}|$, and $1/a^2$ is manifestly independent of $N$.
\section{Results}
\label{sec:the delay function}

For a light emitter that ends up in the same state after each detected emission event, the Fisher information can be computed directly from the waiting time distribution between detection events, \eqref{a}. The waiting time distribution $w(\tau)\mathrm{d}\tau$, in turn, factors into the probability that since the latest detection event, no detection occurred so far, and the conditional probability that the next event happens in the ensuing time interval $\mathrm{d}\tau$, i.e.
\begin{align}
w(\tau)\mathrm{d}\tau &=P(\text{no click in} [0,\tau])\nonumber \\
 &\quad\times P(\text{click in }[\tau,\tau+\mathrm{d}\tau]|\text{no click in} [0,\tau]).
\label{w_P}
\end{align}
Both factors are deducible from a quantum optical trajectory analysis of the photon emission process.

\subsection{Unit detector efficiency; waiting time distribution and Fisher information}

For the two-level atom, the first factor in \eqref{w_P} is given by the trace of the un-normalized "no-jump" density matrix $\tilde{\rho}(\tau)$. This is found by solving the Lindblad master equation \eqref{masterligning_middel} from an initial ground state density matrix, but omitting the term feeding the ground state,
\begin{align}
\mathrm{d}\tilde{\rho}=-i[\hat{H}_0,\tilde{\rho}]\mathrm{d}t-\frac{\Gamma}{2}\lbrace\hat{\s}^{\dagger}\hat{\s},\tilde{\rho}\rbrace\mathrm{d}t.
\label{no jump}
\end{align}
Conditioned on no detection, the state is given by the re-normalized density matrix $\rho(\tau) = \tilde{\rho}(\tau)/\textrm{Tr}\tilde{\rho}(\tau)$. The conditioned decay probability within the next time interval is $\Gamma \rho_{ee} d\tau$. We thus obtain
\begin{align}
w(\tau)\mathrm{d}\tau=\textrm{Tr}\tilde{\rho}(\tau) \frac{\Gamma \tilde{\rho}_{ee}(\tau)\mathrm{d}\tau}{\textrm{Tr}\tilde{\rho}(\tau)} = \Gamma \tilde{\rho}_{ee}(\tau)\mathrm{d}\tau,
\label{delay_rho_ee}
\end{align}
where $\tilde{\rho}_{ee}$ is the $ee$ matrix element of the \textit{un-normalized} density matrix solution to \eqref{no jump}.

The solution of the no-jump master equation, \eqref{no jump}, can be obtained by evolving a pure state vector with the non-hermitian Hamiltonian $\hat{H}_{\textrm{eff}} = \hat{H}_0 - i\frac{\Gamma}{2}\hat{\s}^{\dagger}\hat{\s}$. This evolution is readily solved analytically \cite{CARMICHAEL}. With the system in the ground state at $\tau=0$ and a resonant coupling, $\delta=0$, this yields for the no-jump excited state amplitude:
\begin{align}
\tilde{\rho}_{ee}(\tau)=\left(\frac{\Omega}{2\lambda}\right)^2\sin^2(\lambda \tau) \mathrm{e} ^{-\Gamma\tau/2},
\label{rho_ee_nojump}
\end{align}
where
\begin{align}
\lambda=\frac{\sqrt{\Omega^2-(\Gamma/2)^2}}{2}.
\end{align}
The delay function given by $w(\tau)=\Gamma\tilde{\rho}_{ee}(\tau)$ is presented as the red curve in \fref{fig:delay_test} together with the simulated series of delay times with the same laser-atom parameters.

With a delay function on this form, the integral in \eqref{specialcase} can be performed analytically and it yields the following simple expression for the Fisher information:
\begin{align}
F(\Omega)=N\left(\frac{8}{\Gamma^2}+\frac{4}{\Omega^2}\right) = \frac{4T}{\Gamma},
\label{fisher_analytisk}
\end{align}
where $T$ is the data acquisition time, and we have used that, asymptotically, $N/T$ is given by the mean photon scattering rate, $\Gamma\rho_{ee}^{st}= \frac{\Gamma\Omega^2/4}{\Omega^2/2 + \Gamma^2/4}$ (on resonance). It is remarkable that we obtain such a simple expression, which readily confirms that while the mean scattering rate saturates for strong driving, $\Omega \gg \Gamma$, the sensitivity per detected photon becomes constant, and we can resolve large Rabi frequencies as accurately as intermediate ones. The expression for the Fisher information per detected photon diverges for small Rabi frequencies, The scattering rate, however, depends quadratically on small values of $\Omega$, so per time, the accumulated Fisher information is finite and even independent of the Rabi frequency when the two-level system is driven on resonance.
\subsection{Finite detector efficiency}

It is easy to incorporate the effect of finite detector efficiency $0 <\eta \leq 1$ in the calculation of waiting time distributions: The information retrieved by detection of light with an efficiency $\eta$ is equivalent to the one held by an observer, who is told with a probability $\eta$, whenever the detector clicks in a perfect experiment. When a detector click is reported such an observer knows with certainty that the atom is in the ground state (the "quantum jump" occurs), while in the absence of reported clicks she cannot be certain that no photons were actually detected. This uncertainty is incorporated in the formalism by splitting the feeding term $\Gamma\hat{\s}{\rho}\hat{\s}^{\dagger}$ in the master equation (1) into $\eta \Gamma\hat{\s}{\rho}\hat{\s}^{\dagger}$, representing the reported detection events, and $(1-\eta)\Gamma\hat{\s}{\rho}\hat{\s}^{\dagger}$, representing events, which are not reported and which thus constitute a part of the "no-jump" quantum trajectory dynamics.
Consequently, the conditional no-jump master equation  \eqref{no jump} is replaced by
\begin{align}
\mathrm{d}\tilde{\rho}=-i[\hat{H}_0,\tilde{\rho}]\mathrm{d}t
-\frac{\Gamma}{2}\{\hat{\s}^{\dagger}\hat{\s},\tilde{\rho}\}\mathrm{d}t
+(1-\eta)\Gamma\hat{\s}\tilde{\rho}\hat{\s}^{\dagger}\mathrm{d}t,
\label{master_effektivitet_nojump}
\end{align}
Eq. (\ref{master_effektivitet_nojump}) attains the form of Eq.(1) and Eq.(\ref{no jump}) for $\eta=0$ and $\eta=1$, respectively, and the solution of (\ref{master_effektivitet_nojump}) yields the waiting time distribution for events observed by a detector with efficiency $\eta$,
\begin{align}
w(\tau)\mathrm{d}\tau=\eta\Gamma\tilde{\rho}_{ee}(\tau)\mathrm{d}\tau.
\label{eta_delay}
\end{align}

\begin{figure}
\centering
\includegraphics[width=1\columnwidth]{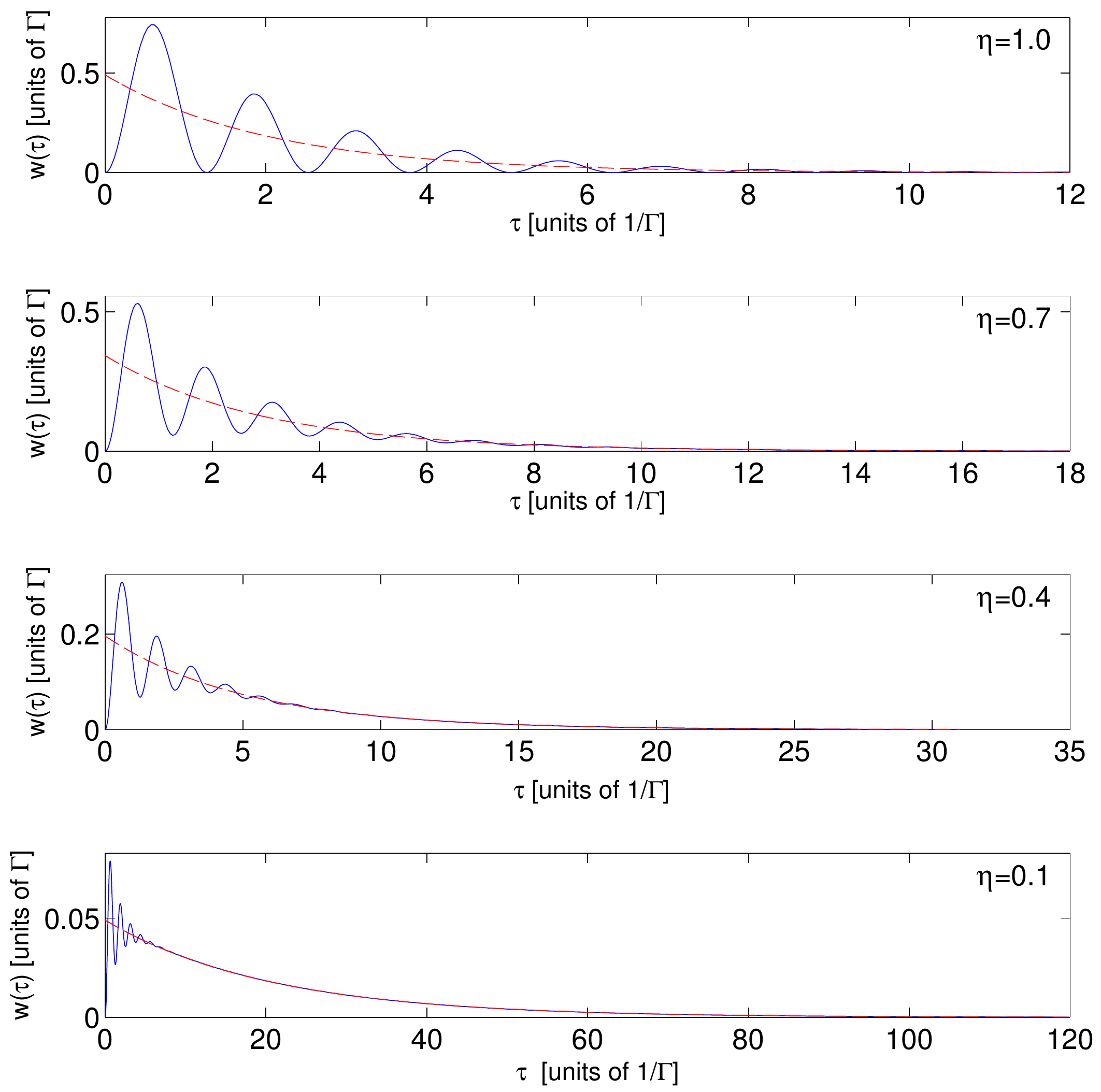}
\caption{\textsl{(Color online) The blue curves show the delay function \eqref{eta_delay} for the two-level atom, calculated for $\delta=0$,  $\Omega=5\Gamma$, and for different values, $\eta= 1,\ 0.7,\ 0.4$, and $0.1$, of the detector efficiency.
The waiting time distributions approach exponential functions for long times \eqref{w_exp}, dashed, red curves in the figures), when the detector is imperfect. Notice that due to the missed detection events, the waiting time distribution extends over longer time when $\eta$ decreases.}}
\label{fig:laveta_delay} 
\end{figure}

The no-jump master equation, \eqref{master_effektivitet_nojump}, constitutes four coupled differential equations, and for arbitrary values of the detection efficiency we have recourse to a numerical solution from which we obtain the waiting time distribution \eqref{eta_delay}. Some characteristic results are summarized in \fref{fig:laveta_delay}. All curves show probability densities and are normalized to unity, but the time scale for the first detection increases when the detector efficiency is reduced. Due to the possibility of missed earlier events the exact nodes in the waiting time distribution for perfect detection disappear, and while a detection within the first few $1/\Gamma$ is likely to report the first actual emission event by the atom, a later detection is almost certainly preceded by unobserved emission of photons by the atom. This explains why the modulation in the waiting time distribution is maintained for short times and gradually replaced by a smooth exponential curve for long times.
Since even small but finite $\eta$ yields a finite fraction of the detection events at short times where the delay function is strongly modulated in time by the Rabi frequency, the sensitivity to the value of $\Omega$ is still significantly improved by considering the actual waiting times rather than only the mean signal.

\begin{figure}
\centering
\includegraphics[width=1\columnwidth]{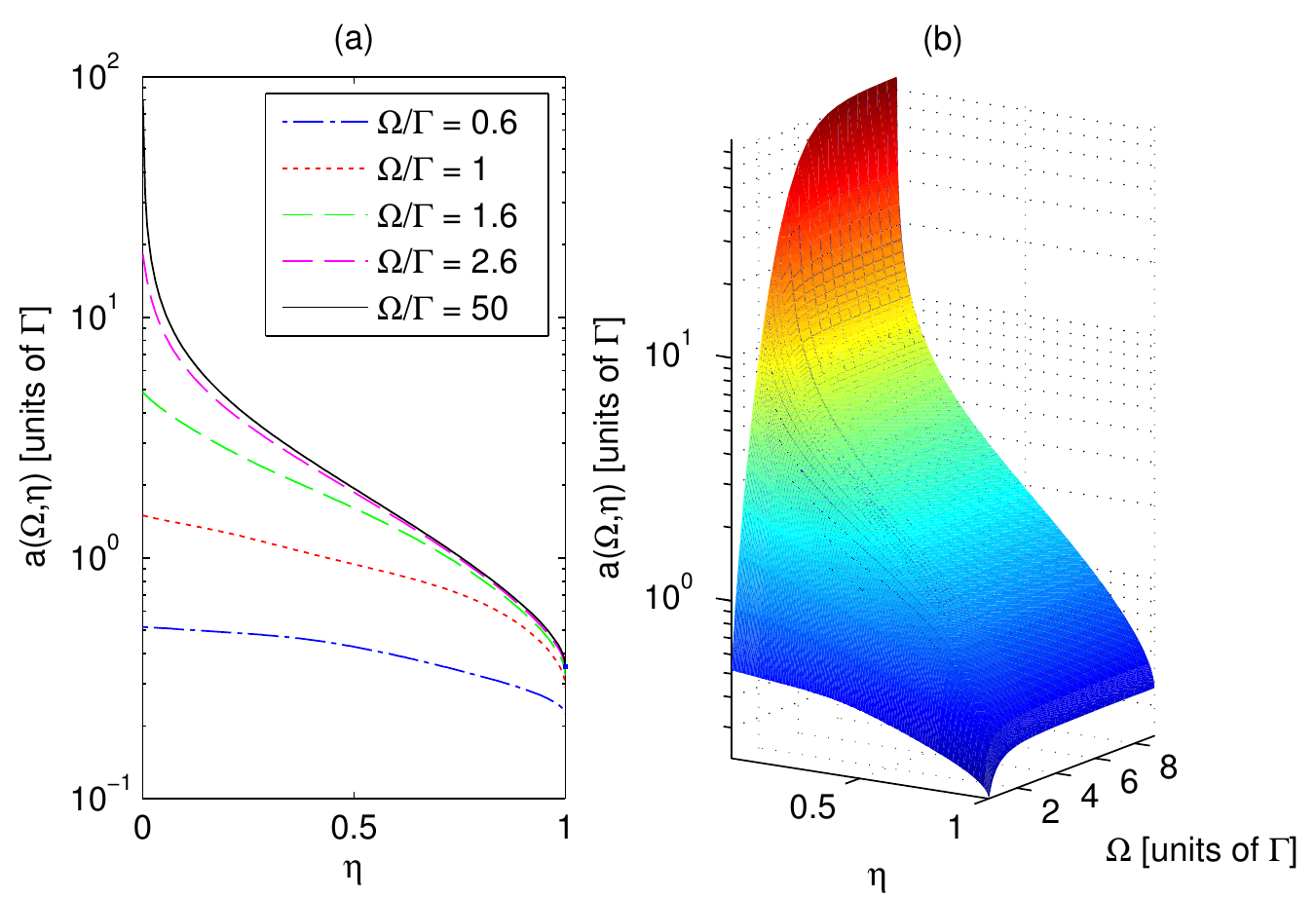}
\caption{\textsl{(Color online) The uncertainty, scaled by $\sqrt{N}$, $a(\Omega,\eta)=\Delta S(\Omega,\eta)\sqrt{N}$, in estimating the Rabi frequency $\Omega$ for a two-level system, driven on resonance ($\delta=0$). The left panel shows the dependence on the detector efficiency $\eta$ for different Rabi
frequencies. The right panel displays the smooth dependence of $a(\Omega,\eta)$ on both parameters $\Omega$ and $\eta$. }}
\label{fig:fisher_effektivitet}
\end{figure}

\fref{fig:fisher_effektivitet} shows $a(\theta,\eta)$ as provided in \eqref{a} and calculated using the waiting time distribution in \eqref{eta_delay} against the detector efficency and laser Rabi frequency. This represents the Cramér-Rao sensitivity bound scaled by $\sqrt{N}$ to make it independent of the photon count.
The $\eta\rightarrow 1$ limit is readily understood through \eqref{fisher_analytisk}. The information content in each detection event saturates at large Rabi frequencies.
We recall  that the Fisher information is given in \eqref{fischer delay} as a factor multiplying the number of detected photons, and that a reduction of the detector efficiency reduces the value $N$ acquired in a given time interval $T$. If we give the Fisher information as a factor multiplying $T$ or, equivalently, the estimate error as a factor multiplying $1/\sqrt{T}$, the $\eta$-dependence therefore becomes even stronger than in the curves and the surface plot in \fref{fig:fisher_effektivitet}.

\subsection{Waiting times for infinitesimal detector efficiency}

In the limit of very low detector efficiency, the detection events occur at a constant rate, $\sim \eta\Gamma \rho_{ee}^{st}$. Their number is Poisson distributed with $\Delta N=\sqrt{N}$. This implies an uncertainty on the Rabi frequency, if it is estimated from $N$, given by $\Delta \Omega = (\partial \Omega/\partial N)\sqrt{N}$, i.e.
\begin{align}
\Delta \Omega \sqrt{N}=\frac{N}{\partial N/\partial \Omega}=\frac{\rho_{ee}^{st}}{\partial \rho_{ee}^{st}/\partial \Omega}.
\label{simpel_eta}
\end{align}
We will now argue that the delay function analysis yields the same result in the limit of low detector efficiency. The no-jump master equation \eqref{master_effektivitet_nojump} approaches the unconditional master equation \eqref{masterligning_middel} when $\eta \ll 1$.
A large majority of the photons are likely to be emitted without detection, so at the first detector click the no-jump master equation has reached the conventional steady state, $=\rho_{ee}^{st}$, which implies that the distribution of delay times is dominated by the exponential tail, indicated also in \fref{fig:laveta_delay},
\begin{align}
w(\tau,\Omega)=\eta\Gamma\rho_{ee}^{st}\mathrm{e}^{-\eta\Gamma\rho_{ee}^{st}\tau}.
\label{w_exp}
\end{align}

The simple expression  \eqref{w_exp} for $w(\tau)$ can be exploited to calculate the Fisher information through \lref{fischer delay}. One readily obtains
\begin{align} \label{lowF}
F(\Omega)
&=N\left(\frac{\partial \rho_{ee}^{st}/\partial \Omega}{\rho_{ee}^{st}}\right)^2.
\end{align}
Finally, through \lref{CRB} this result provides the exact same uncertainty, \eqref{simpel_eta}, as obtained by assuming Poissonian counting statistics.
\eqref{simpel_eta} describes the limit when $\eta\rightarrow 0$ in the plots of \fref{fig:fisher_effektivitet}.

\subsection{Achieving the Cram\'{e}r Rao bound}

We have established that the distribution of waiting times exhausts the information available in the detection record, and that the number of registered occurrences of waiting times in any short time interval is Poisson distributed. It has been shown \cite{CRB}, that for Poisson distributed data records, the CRB can be asymptotically reached by a simple linear estimator
\begin{align}
S(\theta,n(\tau))=\int g(\tau)n(\tau) \, \mathrm{d}\tau + C,
\label{estimator_generel}
\end{align}
which weighs the recorded distribution, $n(\tau) = \sum_i \delta(\tau-\tau_i)$ with an appropriate gain function $g(\tau)$ and adjusts the reference value of the outcome by a constant $C$. We assume that by using a small fraction of the data, we have determined the value of $\theta$ to within a small error $\delta \theta$ which we now wish to reduce by the linear estimator. \eqref{estimator_generel} provides an unbiased estimator for $\delta \theta$ with the choice
\begin{align}
C=-N \int g(\tau)w(\tau,\theta) \, \mathrm{d}\tau.
\label{C_udtryk}
\end{align}
As further shown in \cite{CRB} the variance on the resulting estimator is minimized, and the CRB is indeed reached with the Fisher information in \eqref{specialcase}, if the gain function is chosen as
\begin{align}\label{ggain}
g(\tau)=\frac{\beta}{\Phi(\tau,\theta)}\left.\pdiff{\Phi\ttt}{\theta}\right.
\end{align}
where $\beta=[2N\int\left(\pdiff{\Phi\ttt}{\theta}\right)^2 \mathrm{d}\tau]^{-1}$. Data in complete accordance with the delay function $w(\tau,\theta)$ then leads to
$\delta \theta = 0$, i.e. $S(\theta,\bar{n}\ttt)=0$.

Collecting and rewriting leads to an illuminating form of the linear estimator,
\begin{align}
S(\theta,n(\tau))=\frac{1}{F(\theta)/N}\int \pdiff{w\ttt}{\theta}\left(\frac{n(\tau)}{Nw(\tau,\theta)}-1\right) \, \mathrm{d}\tau.
\label{F_ny}
\end{align}
The prior estimate is adjusted according to the discrepancy between the recorded waiting times and those expected from that prior, and we note that the weight of the delay times is maximal at times where the slope of the waiting time distribution function is high, which confirms our intuition for how the curve and the data in \fref{fig:delay_test} are optimally matched. The  Fisher information per photon count appears as a prefactor and reflects that larger adjustments may apply when the uncertainty is large. One should, however, ascertain  that the adjustment is small enough to validate the linear estimator Eqs. (\ref{estimator_generel},\ref{F_ny}).

\begin{figure}
\centering
\includegraphics[width=1\columnwidth]{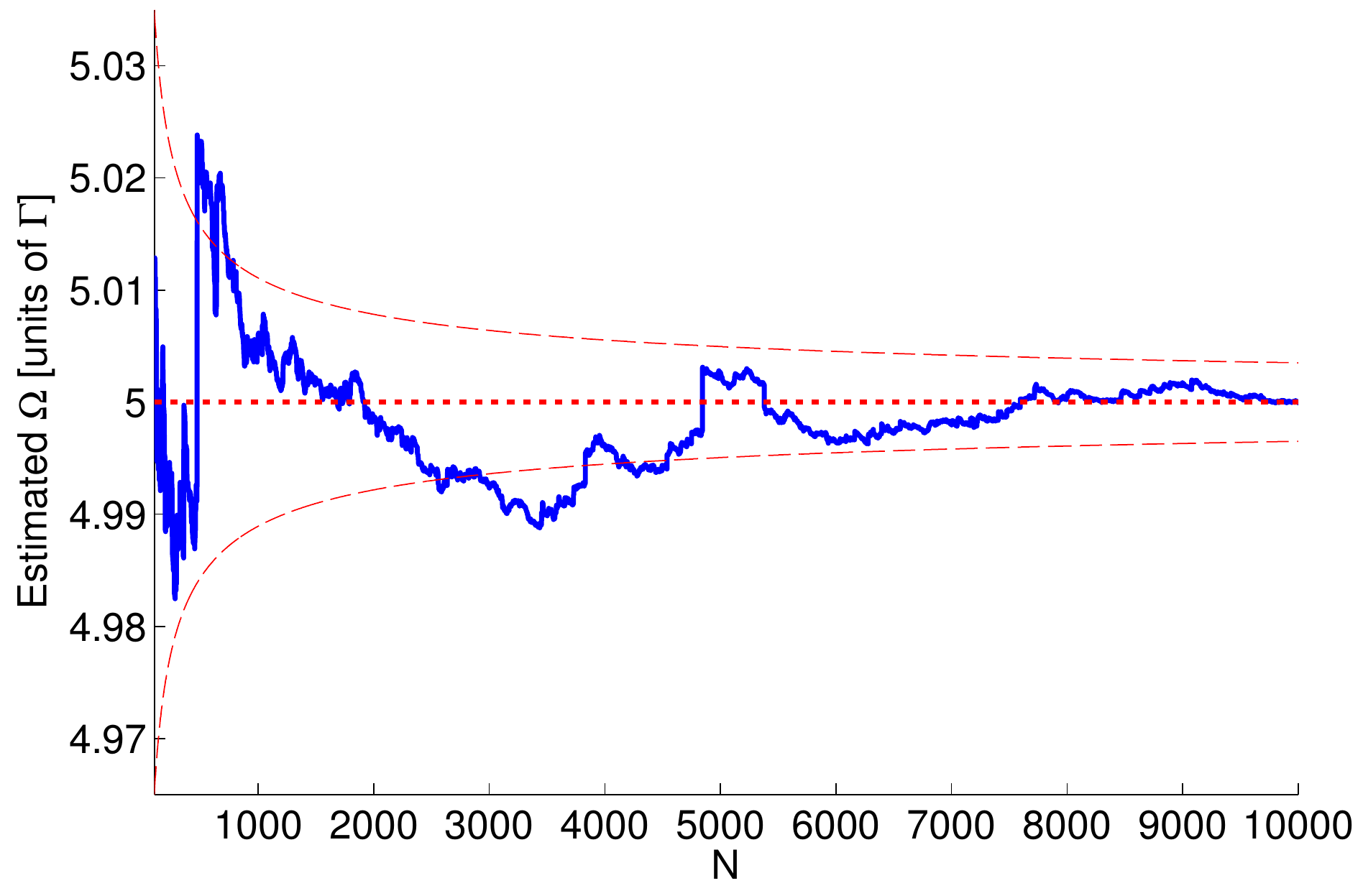}
\caption{\textsl{(Color online) The solid blue line shows the estimate of the Rabi frequency as a function of the number of simulated  photo detection events included in the estimate. The dashed red lines indicate the CRB sensitivity bound, enclosing the actual value (dotted, red line). Results are shown for $100 <N \leq 10000$. The parameter values in the simulation are $\Omega_0=5\Gamma$ and $\delta=0$.}}
\label{fig:esti_N}
\end{figure}

To illustrate the achievements of the estimate \eqref{estimator_generel}, in \fref{fig:esti_N} we show a Rabi frequency estimate (the solid, blue curve) as a function of the number of detection events with unit detector efficiency. The estimate fluctuates around the actual value $\Omega_0=5\Gamma$ (dotted, red horizontal line).
The dashed, red lines in the figure represent $\Omega_0\pm F(\Omega)^{-1/2}$ and show that the deviations of the estimate from the true value are, indeed, compatible with the Cram\'er-Rao Bound in the asymptotic limit.
\vspace{1pt}

\subsection{Estimation of the laser-atom detuning}
While estimation of the Rabi frequency serves as an illustrative example of the Bayesian analysis and the Cram\'er-Rao bound, the analysis applies equally well for the estimation of other system parameters, such as the laser-atom detuning $\delta$. Time and frequency measurements are, indeed, the most precise experiments in chemistry and physics \cite{hall}, and hence a theory for sensitivity limits may be particularly useful for frequencies and detuning parameters. The analysis works in precisely the same way, but the registered delay times should now be matched with the waiting time distribution for different candidate values of the detuning. The solution of the master equation for two-level atom is more complicated when the detuning is varied \cite{torrey}, but the delay function \eqref{delay_rho_ee} can easily be found numerically and applying Eq.(\ref{specialcase}) with $\theta = \delta$ yields the Fisher information associated with the estimation of $\delta$.

 \begin{figure}
\centering
\includegraphics[width=1\columnwidth]{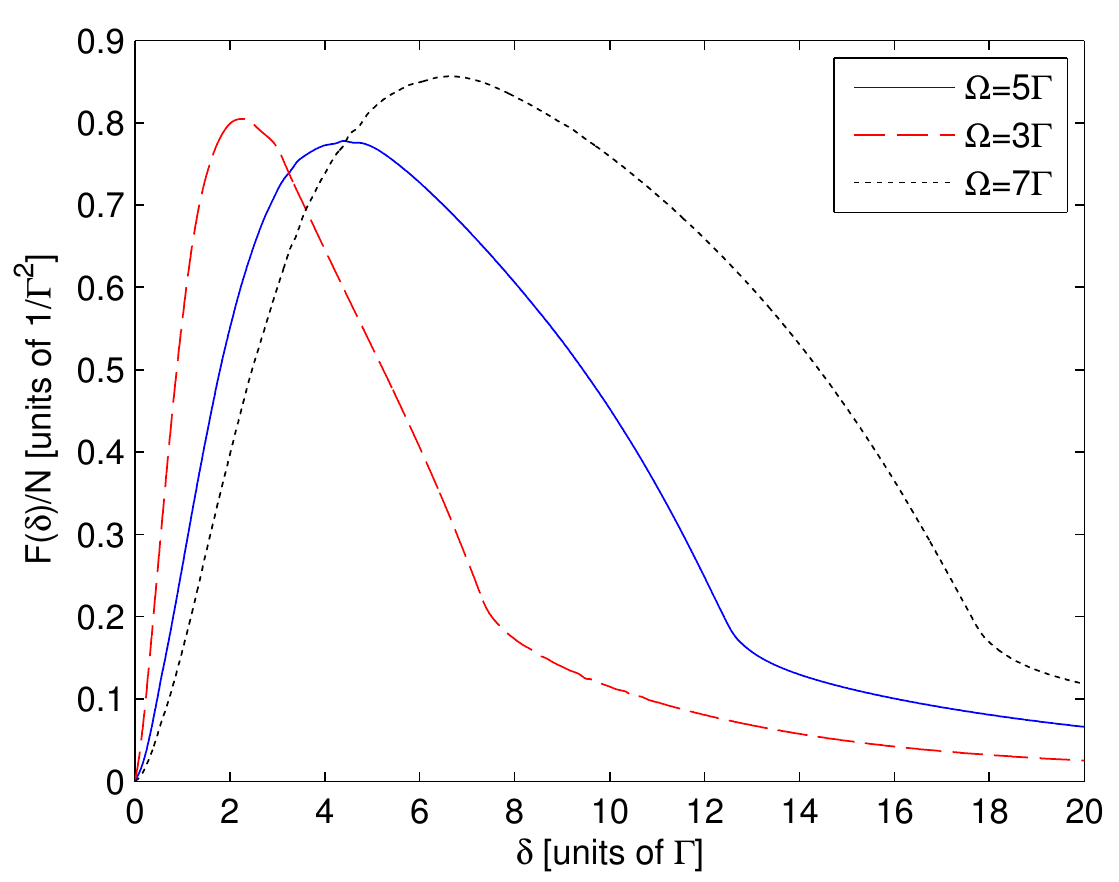}
\caption{\textsl{(Color online) The Fisher information per detection even for estimation of the laser-atom detuning $\delta$ in a two-level system by unit efficiency photon counting. All statistical properties of the counting signal, and hence $F(\delta)$, are even functions of $\delta$. Results are shown for three different values of the Rabi frequency, and it is seen that the information in each detector click is maximized for finite detuning.}}
\label{fig:F_delta}
\end{figure}

\fref{fig:F_delta} shows the Fisher information per detected photon as a function of the detuning for three values of the laser Rabi frequency, assuming unit detector efficiency. The Fisher information is an even function of the detuning. The calculations show that photon counting is not able to discern values of the detuning very close to resonance, and that the sensitivity to variations in the detuning is highest when $\delta$ is in the vicinity of the Rabi frequency. This makes sense, since the fluorescence intensity shows the most pronounced dependence on frequency on the sides of the power broadened line. While this dependence becomes weaker for stronger Rabi frequencies, we recall that the Fisher information pertains to the information about $\delta$ extracted from the waiting time distribution and not only from the mean fluorescence rate.

In \cite{Fisher_open} it is shown that homodyne detection is, indeed, sensitive to the value of $\delta$ around 0. Homodyne detection, however, yields a noisy signal at all times, and is not amenable to the analysis of the present article.

\section{Conclusion and outlook}
\label{sec:conclusion}

In this article, we have presented an analysis of the Fisher information and the Cram\'er Rao sensitivity bound for the determination of the parameters that govern the dynamics of a light emitting system by detection of the emitted radiation. We argued, that for an emitter that decays into the same final state with every emission event, the information available from an entire photon counting signal is fully represented by the distribution of waiting times between the photon detection events. The seemingly very complicated task of determining the resolution limit by an analysis of how the conditional likelihood function varies with the ensemble of typical measurement records, thus reduces to a simple calculation involving the theoretical delay function. An optimal estimate reaching that limit is explicitly provided.

We focused on the example of a two-level system and on the determination of the Rabi frequency, and we included a brief discussion of how the analysis applies for estimates of the laser-atom detuning.  The analysis equally well applies to determine the decay rate of the system, and it may also be applied to more complicated quantum systems as long as every detection event is accompanied by a quantum jump of the system into the same final state. For example, quantum ladder systems with decay from an intermediate state to the ground state are interesting probes, because electric and magnetic fields easily shift the energy of their excited state and significantly influence the system dynamics, e.g., through the mechanism of electromagnetically induced transparency, \cite{PhysRevLett.105.193603}.

Following the general arguments in Sec. III, we have developed theory for systems with several distinguishable decay channels, leading to quantum jumps into different, distinct final states \cite{OnWeGo}. We believe that in addition to their potential use for concrete precision measurements, the two-level systems studied in this article and extended models, tractable by similar means, may provide insights and tests for parameter estimation within a number of more complicated scenarios, including estimation of time dependent parameters and of sets of several unknown parameters.
\\\\
The authors acknowledge discussions with S\o ren Gammelmark and financial support from the Villum Foundation.


%

\end{document}